\documentclass[preprint,showpacs,preprintnumbers,amsmath,amssymb]{revtex4}

\usepackage{epsfig}
\usepackage{graphics}

\begin{document}
\title{Gravitational Waves from Light Cosmic Strings: \\ Backgrounds and Bursts with Large Loops}
\author{Craig J. Hogan}
\address{Astronomy and Physics Departments, 
University of Washington,
Seattle, Washington 98195-1580}
\begin{abstract}
 The mean  spectrum and burst statistics of gravitational waves produced by a cosmological population of cosmic string loops  are estimated using analytic approximations, calibrated with earlier simulations. Formulas are derived showing the dependence of observables  on the string tension $G\mu$, in the regime where newly-formed loops are relatively large, not very much smaller than the horizon.   Large loops form earlier,  are more abundant, and generate a more intense stochastic background and more frequent bursts than assumed in earlier background estimates, enabling    experiments to probe lighter cosmic strings of interest to string theory. Predictions are compared with instrument noise from current and future experiments, and with confusion noise from known astrophysical gravitational wave sources such as stellar and massive black hole binaries.
In these large-loop models, current data from millisecond pulsar timing already suggests that  $G\mu$ is less than about $ 10^{-10}$, close to the minimum value   where bursts might be detected by Advanced LIGO, and a typical value expected in strings from brane inflation.  Because of confusion noise expected from massive black hole binaries, pulsar techniques will not be able to go below about $G\mu\approx 10^{-11}$.   LISA will be sensitive to stochastic backgrounds created by strings as light as $G\mu\approx 10^{-15}$, at frequencies  where it is limited by confusion noise of Galactic stellar populations; however, for those  lightest detectable strings, bursts are rarely detectable. For $G\mu> 10^{-11}$,  the stochastic background  from strings dominates the LISA noise by a large factor, and burst events may also be  detectable by LISA, allowing   detailed study of loop behavior. 
Astrophysical confusion might be low enough at 0.1 to 1 Hz to   eventually reach $G\mu\approx 10^{-20}$ with future interferometer technology.    
\end{abstract}
\maketitle
\section{introduction}
Cosmic strings have been studied for many years as a possible new form of mass-energy with new and distinctive astrophysical effects\cite{kibble,Zeldovich:1980gh,Vilenkin:1981iu,Vilenkin1981,Hindmarsh:1994re,vilenkinshellard} Originally  conceived in  field theory as defects caused by broken U(1) symmetries in Yang-Mills theories,  they have recently re-emerged as  possible structures of  fundamental string theory, called cosmic superstrings,   F-strings and D-strings \cite{Jones:2002cv,Sarangi:2002yt,Jones:2003da,Copeland:2003bj,Jackson:2004zg,polchinski,Polchinski:2004hb,davis,kibble2,vilenkin2} that tend to  arise naturally in models of brane  inflation \cite{Kachru:2003sx}.  In both cases, a cosmic network of  macroscopically extended, quasi-stable strings naturally forms   via the Kibble mechanism, a process of quenching after cosmological inflation. Although strings are now known not to contribute significantly to the formation of cosmic structure,  other  gravitational interactions of even very light strings  may be observable and give   direct data on their fundamental nature. The most novel and conspicuous  effect comes about from their principal decay process:  the primordial network of long strings continually intersects with itself, spawning isolated, oscillating loops that ultimately radiate almost all of their energy into 
gravitational waves\cite{Vilenkin:1981iu,Hogan:1984is,Vachaspati:1986cc,Bennett:1989yp,Caldwell:1991jj,Caldwell:1996en}. 

This paper estimates the   spectrum and burst statistics of the cosmic gravitational wave background produced by loops of light cosmic string, and compares those predictions with capabilities of  current and planned experiments that span almost  a factor of $10^{12}$ in frequency: millisecond pulsar timing, the space interferometer LISA, and ground-based interferometers such as LIGO, VIRGO, and GEO.  
The main new results arise from renewed attention to the effects of the size spectrum of string loops: if a non-negligible fraction of loops stabilize when they are relatively large compared to the horizon, the surviving population forms earlier and loops are more abundant. As a result, the strings generate more gravitational waves, so that even very light strings are detectable.

Although the fundamental physics differs widely for different types of strings,   just two main parameters, depending on that physics,  control  their astrophysical behavior and gravitational effects: the dimensionless mass per length or tension $G\mu$ (in Planck units with $G=m_{Planck}^{-2}$, and setting the speed of light  to unity) and the interchange probability $p$. 
The main aim here is to explore the sensitivity and relative strengths  of the various experiments at small values of   $G\mu$.  
The predictions here for both the noise sources and the string backgrounds are uncertain by at least a factor of a few, but are good enough to estimate the  experimental sensitivity to (and currently allowed range for) string parameters, to identify promising approaches for improvement, and to assess in general terms the science impact of the experimental projects.

In addition,  other parameters retained here, $\gamma$ and $\alpha$, characterize   uncertainty or controversy at present of certain aspects of the string behavior.   The dimensionless decay rate $\gamma$ is generally taken to be around 50 and is  unlikely to differ from this by a large factor. However, 
dynamical simulations  of network behavior disagree about $\alpha$, the size of newly formed loops in units of the Hubble scale $H^{-1}$, and this  strongly affects the predicted backgrounds\cite{Bennett:1989yp,Caldwell:1991jj,Caldwell:1996en,Allen:1990tv,Austin:1993rg,Siemens:2002dj,Siemens:2003ra,Vanchurin:2005pa,Ringeval:2005kr,Martins:2005es,Sakellariadou:2004wq,Avgoustidis:2005nv}.  Recent work\cite{Vanchurin:2005pa}  suggests that 
that    newly formed loops are smaller than the horizon by a factor of the order of $\alpha \approx 0.1$, in contrast to    earlier estimates  of the order of $\alpha\approx G\mu$ or even smaller--- a difference of many orders of magnitude.  The   approximations adopted here are valid over a range of relatively large $\alpha$, $0.1>\alpha>10^{5}\gamma G\mu$, that is consistent with the new simulations designed to accurately estimate $\alpha$\cite{Vanchurin:2005pa},  but that has not been included in recent background estimates.  Within that range, the predicted spectra simply scale as  flux $\propto\alpha^{1/2}$. As shown below, larger  $\alpha$  leads to more loops,  stronger   predicted backgrounds and hence greater sensitivity to light strings than earlier estimates. 

The analytical model of the network behavior, loop formation and loop radiation adopted here shares the basic framework of previous models. The most important difference is the choice   of a large value of $\alpha$, which leads to the approximation that loops dominating gravitational wave production   form during the radiation era instead of the matter-dominated era, and because of the increased intensity,  leads to consideration of  lower  $G\mu$ than earlier studies.    Some new features need to be added in turn because of the  very low  $G\mu$,  since some experiments see the fundamental frequencies ($f=2/L$, where $L$ is the length) of loops that have a lifetime   larger than the Hubble time today.
 Aside from those differences, the model here follows standard ``one scale'' approximations for string behavior. We adopt earlier calculations\cite{Caldwell:1991jj,Caldwell:1996en} as a normalization in high frequency limit, and  scale these results to lower frequencies and lower $G\mu$. 

The rate of observable beamed bursts of radiation from string cusps is also estimated, and roughly agrees with extrapolations of other recent studies\cite{Damour:2000wa,Damour:2001bk,Damour:2004kw,Siemens:2006vk}.  However,   the abundant loops arising from large $\alpha $  lead to a situation where bursts are also accompanied by a more  intense stochastic background. The larger overall flux of stochastic  radiation, especially from high redshift loops, is detectable at very low $G\mu$, and  for the lightest strings detectable via their stochastic background, bursts are rarely detectable, both because the confusion is greater and because the fundamental frequencies are higher. (The detected frequencies of the stochastic background can be close to the fundamental frequencies of the most abundant  loops, where they are not beamed substantially at all.)  On the other hand somewhat heavier strings produce both an easily detectable stochastic background and a possibly detectable rate of bursts, detectable at high frequencies, especially with Advanced  LIGO. In this case the unique identifiable signature of the  bursts provides  independent information about the loop parameters and behavior, and a validation of the source model.

Another major difference from previous work on string detection is a more comprehensive comparison with sources of astrophysical confusion noise limits from populations of binaries. At low frequencies, including those accessible to LISA and to pulsar timing techniques, these will eventually be more important and fundamental than instrument noise in setting the ultimate attainable sensitivity. 
The confusion noise and instrument noise models for LISA frequencies are largely taken from\cite{Hogan:2001jn}. 

Current models of string-based inflation provide some context for the scale of string parameters.  Inflation is an effective theory, not a fundamental one. It describes the behavior and cosmological gravitational effects of a scalar inflaton field in terms of an effective potential; the goal has always been to relate that field and potential, at a deeper level, to other fields including the Standard Model. In string theory,  the recent worked examples with such a physical basis are   models of brane inflation. An effective  inflaton   can be realized where the scalar field represents  the separation of a pair of  D3/${ \hspace{1pt}\overline{\hspace{-.5pt}{\rm D3}\hspace{-1.5pt}}\hspace{1.5pt} }$   branes\cite{Kachru:2003sx}. At the end of inflation, the branes annihilate and 
break  U(1)$\times$U(1) symmetries   giving rise to 
cosmic Dirichlet strings (D1-branes)  and fundamental (F-) strings.  In these models of brane inflation, if parameters are chosen to create the observed cosmic perturbations ($\delta \phi\approx 10^{-5}$)  via the usual scalar quantum fluctuations,  a  wide range of string tensions is possible,  $10^{-12}<G\mu<10^{-6}$\cite{Sarangi:2002yt,Jones:2003da,
polchinski}, with a typical value   $G\mu\approx 10^{-10} $.
The range of $p$ is not known but a variety of models yields a typical range $10^{-3}<p<1$\cite{Jackson:2004zg}.
The analysis here shows that current and planned gravitational-wave experiments probe   this entire range of parameters and well beyond. Indeed, apart from the remaining doubts about $\alpha$, the arguments above suggest that the typical value of $G\mu\approx 10^{-10}$ is already in conflict with pulsar timing data.
From this perspective, current and future gravitational wave experiments contribute unique and useful   information about  inflation and its relation to new, stringy physics.

\section{model of string loop populations and radiation spectrum}

Many aspects of  the behavior  of strings in a cosmological context are well established \cite{kibble,Zeldovich:1980gh,Vilenkin:1981iu,Vilenkin1981,Hindmarsh:1994re,vilenkinshellard}.
The Kibble mechanism lays down local gauge strings initially as a set of nearly random walks with a very small comoving coherence length. These then evolve obeying the Nambu-Goto action; in an appropriate gauge, the local equation of motion of the strings is just the relativistic wave equation.
  On scales larger than the Hubble length $H^{-1}$, the network is ``frozen'' and stretches with the cosmic expansion.  On smaller scales, the strings move with velocities of order $c$. They occasionally cross and exchange partners, with probability $p$ at each crossing. In this way, closed loops separate from the network of long strings. 
 It is this population of loops that dominates the production of gravitational waves.
   Each loop oscillates at a set of discrete set of frequencies $2n/L$ depending on its length $L$, and loses energy slowly due to gravitational radiation. We assume here that gravitational radiation  is the main energy loss, which is the case for many varieties of fundamental strings.   
  
  We adopt the familiar ``one-scale'' model of the loop population \cite{Hogan:1984is,Vachaspati:1986cc,Bennett:1989yp,Caldwell:1991jj,Caldwell:1996en}. 
  The typical size of a newly born loop is parameterized conventionally as $\alpha/H$. The fraction of  total horizon mass in such loops is of order $G\mu$, times a numerical  factor of order unity we take from numerical calculations of the scale-free, high redshift solution. The gravitational wave background at each frequency is a sum of contributions from loops at all redshifts contributing at that frequency; here we  approximate this by estimating the redshifted radiation from the brightest loops at each redshift.  As we see below,  the loop contributions fall naturally into  two main populations: a high-redshift H population of long-decayed loops, radiating from close to the fundamental mode,   and  redshifted by a large factor; and a present-day P population, where the redshift is of order unity or less.  The H  loops that decayed at high redshift dominate the integrated flux at high frequency, in the form of a nearly Gaussian stochastic background. 
  The P loops contribute significantly near  their fundamental frequency,  as well as  a power law spectrum $\propto f^{-1/3}$  above it; they dominate the low frequency background, and are also the source of rare brief burst events at high frequency.  (Although this nomenclature is useful for discussion,  one should bear in mind that the H and P contributions are comparable, within a factor of a few in total flux, over a range of about a thousand in frequency.) 
    
  A population of loops of a variety of sizes and redshifts produces a background with a nearly continuous spectrum. At any given time, the power in a broad band peaks near a frequency around the fundamental frequency of loops whose decay rate is $H$,
  \begin{equation}
  f_{peak}= 2H/\gamma G\mu 
     \end{equation}
  where $\gamma\equiv 50\gamma_{50}$ represents  the dimensionless gravitational radiation power  in all modes\cite{Caldwell:1991jj}, estimated to be $\gamma_{50}\approx 1$.
  At present, the peak of the spectrum is at about
   \begin{equation}
  f_{peak}\approx 10^{-10}(\gamma_{50} G\mu/10^{-9} )^{-1}{\rm Hz}.
 \end{equation}
   At frequencies near $f_{peak}$, the background comes mainly from the present-day population P.     
 
 We assume here that    loops formed at high redshift, during the radiation era. This is true at all  the frequencies we are concerned with,  as long as the P population loops formed before matter-radiation equality at  $t_{eq}$, which requires:
 \begin{equation}
 \gamma G\mu/H_0<\alpha/H_{eq},
 \end{equation}
 or inserting numbers from concordance cosmology (see below),
 \begin{equation}
 \alpha> 10^5\gamma G\mu.
 \end{equation}
 Thus the approximations used here can only be applied consistently to light strings and relatively large $\alpha$.

The intensity of the background in this situation is determined by the production rate of loops in the radiation era. Indeed, at sufficiently high frequencies, the observed radiation comes from loops which both form and decay in the radiation era.  During the period  between the formation of a loop and the time it decays, its energy does not  redshift; thus the  ratio of gravitational wave energy to radiation energy density is of order $G\mu (Z_{form}/Z_{decay})p^{-1} $, where $Z_{form}$ and $Z_{decay}$ denote the inverse scale factors for formation and decay of typical loops, and the mass of newly formed loops is about a fraction $G\mu$ of the total horizon mass.
The spectrum in this limit is flat in energy density, and does not depend on frequency\cite{Vilenkin1981,Hogan:1984is}:
 \begin{equation}
 \Omega_{GH}  =[\gamma_{50} G\mu\alpha]^{1/2}p^{-1}\Omega_R F_L
 =10^{-8}(G\mu/10^{-9})^{1/2}p^{-1}\gamma_{50}^{1/2}\alpha_{0.1}^{1/2},
  \end{equation}
where $\Omega_{GH}$  refers to the energy density in gravitational waves   per $\log f$ in the high frequency limit, in  units of the critical density, and $ \Omega_R$ denotes the total energy density in relativistic species. The coefficient $F_L $ is a numerical factor of order unity characterizing the creation rate of new loop mass per Hubble time in a scaling solution; its value only depends on the mean equation of state at loop formation being that of relativistic particles ($w=1/3$). The factor used here is taken so that the spectrum normalization agrees with  numerical estimates of \cite{Caldwell:1991jj} in the high $f$ limit, and this is   the normalization for all the spectra estimated here. A family of these H spectra is shown in Fig. 1 for various values of $G\mu$.

The scaling $p^{-1}$ comes from a very simple argument: to achieve a scaling solution in a one-scale model, each long string must   spawn loops a rate independent of $p$, therefore the number of long strings it needs  to collide with varies like $p^{-1}$. This scaling  has been questioned in  detail\cite{Avgoustidis:2005nv,Sakellariadou:2004wq}   but we adopt it here as a simple placeholder.

This flat  spectrum holds approximately down to the frequency $f_{eq}$ of waves coming from  loops that are decaying at the epoch of matter-radiation equality. In concordance cosmology we have $Z_{eq}\equiv 1+z_{eq}\approx \Omega_M/\Omega_R\approx 3300$, and $H_0t_{eq}\approx 10^{-5}$; thus 
\begin{equation}
f_{eq}/f_{peak}=(H_0t_{eq}Z_{eq})^{-1}\approx 30.
\end{equation}

For $f_{eq}<f<f_{peak}$, the spectrum from the loops decaying at $Z_{decay}(f)$ increases above $\Omega_{GH}$. The typical loops decaying during matter domination, but before $Z\approx 2$, during which $Z\propto t^{-2/3}$, scale like $Z_{form}\propto t^{-1/2}$,  $Z_{decay}\propto t^{-2/3}$, and $f\propto Z_{decay}^{-1}t_{decay}^{-1}$. Thus they
produce a spectrum of redshifted radiation $ \Omega_G\propto (Z_{form}/Z_{decay}) \propto (f/f_{eq})^{-1/2}$.  With this scaling, the radiation from intermediate redshifts, though not negligible, is however subdominant compared to the high frequency tail from the  P population loops at $Z< 2$, which goes like $\propto f^{-1/3}$; that is why it makes sense to conceptually separate  the  H and P populations for describing the effects.

For this analysis we  neglect features in the spectrum near $f_{peak}$ due to cosmic acceleration since redshift of order 1, which are not neatly accommodated by simple scaling arguments.  Neglecting the brief accelerating phase does not at all affect the estimate of the H population  or  
$f_{eq}<f<f_{z\approx 1}$ contributions to the stochastic background, which are normalized to the microwave background flux.   It does alter   the spectrum near $f_{peak}$ and reduce the contribution of the P population, since the older universe at given $H_0$ means currently-decaying loops of size $L_{decay}$ are slightly larger (hence rarer, relative to the radiation density) than in the matter-dominated scaling.  The density of nearby loops is pegged to the (model-independent) radiation density, times a factor $Z_{form}/Z_{decay}$; this factor is proportional to the $L_{decay}^{-1/2}$; since the accelerating universe is about twice as old as the matter-dominated model of the same $H_0$, the frequency at $L_{decay}$ is about half as large, their mass density is smaller  by about ${2^{-1/2}}$, and their gravitational-wave energy flux is smaller by about $2^{-3/2}$.  Thus the numbers below overestimate the   burst rate from nearby loops and the  mean flux at $f<f_{peak}$ from those loops, by a factor of a  few. Similar factors contribute to some of the differences between previous estimates of burst rate, for example between \cite{Damour:2004kw} and \cite{Siemens:2006vk}; the present approximation is closer to \cite{Damour:2004kw}. The estimates of stochastic background flux  above $f_{peak}$ are little affected by this approximation. 

Since loops do not radiate below their fundamental frequency,  at  low  frequencies $f<f_{peak}$ the background comes from loops whose lifetime exceeds $H_0^{-1}$.  The energy density  spectrum is given by the mass density $\Omega_L(f^{-1})$
of loops of size $L\approx 2f^{-1}$, which in turn is $\propto Z_{form}^{1}\propto L^{-1/2}\propto f^{1/2}$,  times the fraction of their rest mass they radiate in a Hubble time, which in turn is $\propto L^{-1}\propto f$:
\begin{equation}
\Omega_G\propto \Omega_L(f^{-1})(f/f_{peak})\propto f^{3/2}.
\end{equation}
This regime was often not considered in earlier treatments because unless $G\mu$ is very small, these fundamental  frequencies are too low to be observed directly.  Although the normalization of this contribution is overestimated here due to neglect of recent acceleration, the slope should be accurate.
The $\Omega_G \propto f^{3/2}$ scaling applies only down to the frequency corresponding to loops formed at $t_{eq}$, which have $f\approx 2.5\times 10^{-12}\alpha_{0.1}^{-1}$Hz. 

 Summarizing the above contributions, the spectrum is:
\begin{eqnarray}
\Omega_G(f) & \approx & \Omega_{GH} [1+\sqrt{30}(f/f_{peak})^{-1/3}],\qquad  f >30 f_{peak}  \\ \nonumber
& &  \Omega_{GH}[\sqrt{30}(f/f_{peak})^{-1/2}+\sqrt{30}(f/f_{peak})^{-1/3}],\qquad  \qquad  f_{peak}<f <30 f_{peak}  \\ \nonumber
& &  \Omega_{GH}\sqrt{30}(f/f_{peak})^{3/2},\qquad  \qquad  2.5\times 10^{-12}\alpha_{0.1}^{-1}{\rm Hz}<f < f_{peak}.  
\end{eqnarray}
Spectra in Fig. 1   show contributions from the various populations at each frequency. The radiation in these models is much more intense than estimates based on smaller values of $\alpha$, where the present-day population is made of loops which formed during the matter era. As a result, the current and future sensitivity to $G\mu$ is correspondingly greater. 

We have adopted the conventional size parameter $\alpha$ for the one-scale model, but  in reality not just the characteristic size, but also  the detailed shape of the initial size spectrum matters. 
For such small $G\mu$, even   if  almost all loops shatter into tiny pieces quickly,  a small fraction of loops that stabilizes close to the horizon scale  can still come to dominate the   loop population. The fraction of horizon-size loops that needs to stabilize in order for radiation-era loops to dominate near $f_{peak}$ is only  $\approx (G\mu)^{1/2} (\Omega_M/\sqrt{30}\Omega_R)\approx 600 (G\mu)^{1/2}$ (times some numerical factors), which is a small number for the light strings we are contemplating.
 Unless $\alpha$ is relatively large,  accurate estimates of the radiation background will require estimates of loop spectra with very large dynamic range.

\begin{figure} \label{fig: figure1}
\epsfysize=3.5in 
\epsfbox{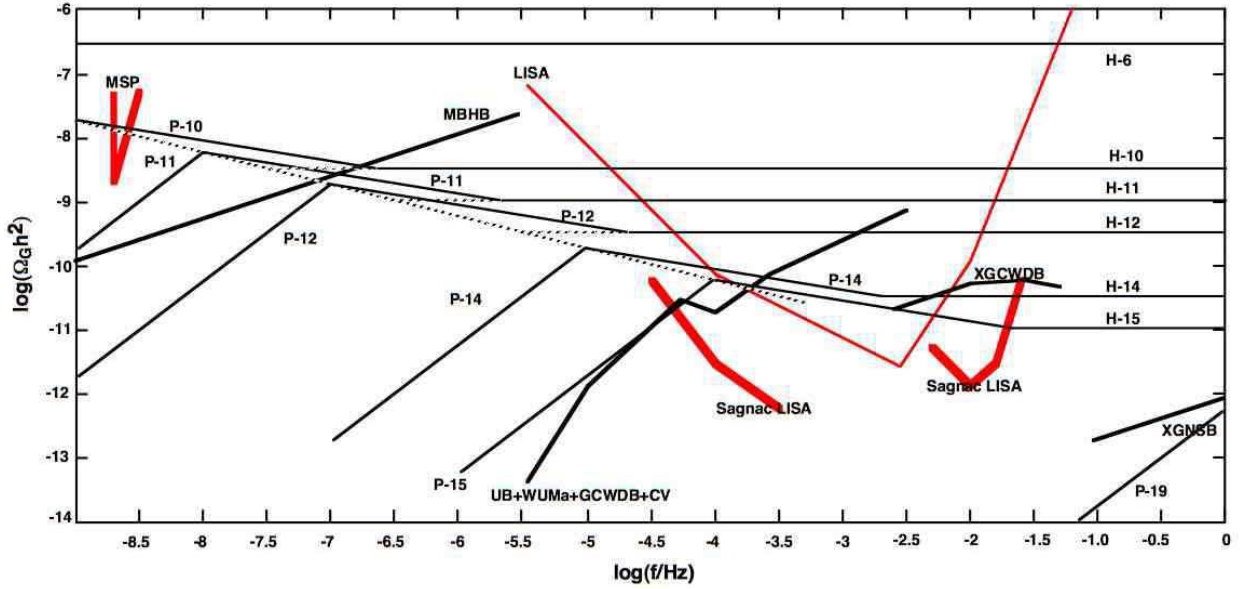} 
\caption{Predicted gravitational wave backgrounds from strings,  and  noise sources. Broad band energy density is shown in units of the critical density for $h_0=1$, as a function of frequency, for $\alpha=0.1$. Noise levels are shown for current millisecond pulsar data (MSP), and projected  LISA sensitivity in maximum resolution and Sagnac modes. Confusion noise is shown for massive black hole binaries (MBHB), the summed Galactic binary population including binary white dwarfs (UB+WUMa+GCWDB+CV), and extragalactic populations of white dwarfs (XGCWDB) and neutron stars (XGNSB). Radiation from   loop populations at  high redshift (H)  and present-day (P) is shown, labled by the value of $G\mu$. Dotted curves show the  contributions of $z>1$ loops where they are subdominant to the P contributions. Current (MSP) sensitivity is at about $G\mu\approx 10^{-10}$, and LISA will reach to around $G\mu\approx 10^{-15}$.}
\end{figure} 
\break

\section{burst detection and self-confusion noise}

At  frequencies  $f  $ much higher  than the fundamental mode, the radiation is increasingly more highly beamed and episodic. 
The most important characteristic of loop radiation far above the fundamental frequency arises from the catastrophes called cusps.\cite{Turok:1984cn}  
It can happen that in the course of a typical loop orbit,  at some instants,  at single points on the world-sheet, the velocity of the string is formally the speed of light. The vectors describing the derivatives of the string trajectories can be mapped onto  curves on a unit sphere for left- and right-moving modes, and cusps occur at some point whenever these intersect. Since curves ``typically'' intersect on a sphere,  it  is thought  that the behavior happens reasonably often, of order once per orbit for a typical loop. 

The behavior of cusps   obeys simple, universal scaling deriving just from the geometry of the events.
The metric strain obeys  a universal, power-law waveform that includes a power-law tail of energy up to high $f$.  Approaching time $\delta t$ from the moment of a typical catastrophe,  in units  given by the fundamental mode of the loop, the metric strain amplitude due to  radiation beamed from a cusp varies like  $h\propto \delta t^{1/3}$, and is beamed within an angle $\theta\propto\delta t^{1/3}$.  (That is, if observed at angle $\theta$, the cusp behavior is smoothed out at $\delta t< \theta^3$.) The time-integrated  spectrum at $f$ is dominated by the waveform at $\delta t\approx f^{-1}$. Thus the integrated energy   is $\propto h^2 f^2\theta^2\delta t\propto f^{-1/3}$.

These events are called ``bursts''. (Even   though the amount of energy radiated close  to a cusp   actually decreases as $\delta t^{1/3}$, the power within the beam, in a particular direction, does increase like $\delta t^{-2/3}$ until the smoothing sets in.)  The distinctive signature of a single burst  stands out from other astrophysical sources, and can be distinguished if it has enough power relative to other noise sources, including the confusion noise of other loops and distant bursts. As discussed below, even with the optimistic assumptions made here, bursts are not in general the most easily detectable feature of very light strings; on the other hand, they can  be a powerful diagnostic even if they are fairly rare.

The number of loops contributing to the background from  population H is so large that its contribution is for all practical purposes gaussian. The number of loops contributing  from  population P is also large near $f_{peak}$ but decreases at higher frequency since the radiation from each loop is concentrated in angle and  in time.  Among the P loops,  some   are much closer than the mean so they can occasionally be bright enough to poke above the background and  produce a detectable burst.  We want to estimate the rate of the detectable bursts. 

The rate of burst events was estimated in the important papers by Damour and Vilenkin\cite{Damour:2000wa,Damour:2001bk,Damour:2004kw}. A more detailed analysis by Siemens et al. \cite{Siemens:2006vk} yields 
lower rates due to different (and more realistic) cosmology and more demanding  criteria for detection. The rates estimated here are not as reliable as Siemens et al.  and err on the side of large burst rates.  Even so, we find that bursts are not an   important observable for the very lightest strings, because they are so light that they are observed rather close to their fundamental mode.
 We   make several  simplifying   assumptions, such that all the radiation in the high-frequency tail is from cusps and that there is one cusp per orbit. This makes no difference in estimating the strength of stochastic backgrounds, but it gives an optimistic estimate of  the frequency of detectable bursts.  Event rates simply scale directly with these numbers.

The total number $N$ of loops in the P population in a volume $V$  is about
\begin{equation}
N\approx {\rho_{crit}\Omega_G(f_{peak})V\over  M_L(f_{peak})}
\end{equation}
where $M_L(f_{peak})\approx \mu / f_{peak}$ is the typical mass of one of the loops. Adopting a volume $V=4\pi/3H_0^3$ and the spectrum as estimated above for $\Omega_G(f_{peak})$ yields an estimate of the total number of individual sources contributing near the peak of the spectrum,
\begin{equation}
N(f_{peak}) 
\approx   10^{9} (G\mu/10^{-9})^{-3/2}p^{-1}\gamma_{50}^{-1/2}\alpha_{0.1}^{1/2}.
\end{equation}

Note that in the same volume there are about $10^9$ bright galaxies;  for strings lighter than $G\mu\approx 10^{-9}$, loops  are     more numerous than galaxies.

 It is important for the following that the loops are distributed uniformly in space; we would get a very different answer for burst rates if  they clustered in the halo of the Milky Way like dark matter. But since  the center of mass of  most of the loops   moves rapidly through space, due to the ``rocket effect'' of beamed gravitational radiation,   their large peculiar velocity likely prevents them from easily binding into galaxies\cite{Hogan:1984is}. Maximal beaming predicts that a loop   has velocity of order unity by the time it decays (in spite of cosmological redshifting over a loop's lifetime, since most of  the recoil momentum is imparted coherently over a decay time).  As long as a typical loop has $v>10^{-2.5}$, more than the typical binding velocity of galaxy systems,  a seemingly safe but still unproven assumption, the bulk of them  remain unclustered.  (As an aside, note that if a loop does by chance become bound its acceleration is not typically enough to allow it to leave a galaxy unless by chance the beaming is carefully controlled directionally to match the orbit. Thus   a small  population of loops does become bound to galaxies and clusters. For very light strings, there might be enough bound to our galaxy to affect the estimate of burst rates.)

Near $f_{peak}$, the radiation from each loop is  radiated in a broad beam pattern and more or less continuously. At higher frequencies, the power in the $f^{-1/3}$ tail is beamed in angle and concentrated in time around the cusps. The fraction of loops beaming in our direction is $\propto \theta^2\propto f^{-2/3}$, while the fraction of time radiating in a band around $f$ is $\propto f^{-1}$; therefore the number of loops contributing to the background at any given time from population P at frequency $f>f_{peak}$ is about
\begin{equation}
N(f)\approx N(f_{peak}) (f/f_{peak})^{-5/3}.
\end{equation}

In order to detect a burst with a matched filter with high confidence, the power of the burst must substantially exceed the noise power in the rest of the stochastic background during the burst. Since the burst only lasts at each frequency  for a time $t\approx f^{-1}$,  the power  exceeds the background by a factor of about $SNR^2$ where  $SNR$ is the signal-to-noise ratio in the metric strain; because they are rare events, we will require  $SNR>5$ for a detection, and more for a waveform study.  For $N(f)>>1$, this happens only when a burst occurs unusually close to us, so that it stands out from the background.
 We estimate the mean time between burst events that stand out from this background.

In the first instance,  assume that the only background is from the same loop population that is producing the bursts, that  is, made mostly of bursts from the P loops at distance $\approx H_0^{-1}$. 
In order to be detectable above this background, a burst has to   be within a distance $D$, where $DH_0\approx SNR^{-1} N(f)^{-1/2}$.    The total number of loops (including those not beamed at us) within this distance is then
\begin{equation}
N_D\approx SNR^{-3} N(f_{peak})N(f)^{-3/2}.
\end{equation}
Each oscillation (that is, in a time $f_{peak}^{-1}$), the probability of a given loop sending a burst at us at frequency $f$ at some point  is about $(f/f_{peak})^{-2/3}$. The rate of observable events is then
\begin{equation}
\tau^{-1}\approx N_D f (f/f_{peak})^{-5/3},
\end{equation}
leading to a mean time between observable events,
\begin{equation}
\tau_{P>H}\approx SNR^{-3} N(f_{peak})^{1/2}(f/f_{peak})^{-5/6}f^{-1}
\end{equation}
or
\begin{equation}
\tau_{P>H}\approx
6\times 10^{10}\ {\rm sec}\ \ SNR_5^3(G\mu/10^{-14})^{-19/12}p^{-1/2}(f/10^{-2.5}{\rm Hz})^{-11/6}
\gamma_{50}^{-13/12}\alpha_{0.1}^{1/4},
\end{equation}
where we require $SNR_5\equiv SNR/5$ to be at least of order unity for a significant detection, and significantly higher for a confirmation of the cuspy waveform shape.
We can ask for example whether a burst might be seen by LISA from very light strings. Since  the optimal frequency in this case $f=10^{-2.5}$ Hz, we  conclude that for very light strings near the detection limit, bursts  are likely  to be    rare,  at best  occasionally detectable in an experiment lasting several years. 

Now consider the regime where the bursts have to compete with a dominant H background. In that case, the detection distance $D$ is smaller; it is multiplied  by the square root of the P to H flux ratio, $\approx [30^{1/2}(f/f_{peak})^{-1/3}]^{1/2}$. The time between detectable events is longer by a factor $\propto D^{-3}$, so
\begin{equation}
\tau_{P<H}\approx 8\times 10^{-2} SNR^3 N(f_{peak})^{1/2}(f/f_{peak})^{-1/3}f^{-1},
\end{equation}
or
\begin{equation}
\tau_{P<H}\approx 6\times 10^8 \ {\rm sec}\  SNR_5^3 (G\mu/10^{-12})^{-13/12}p^{-1/2}(f/10^{-2.5}{\rm Hz})^{-4/3}\gamma_{50}^{-7/12}\alpha_{0.1}^{1/4}.
\end{equation}
Thus at around  $G\mu\approx 10^{-11}$ and above,
 burst events   are common enough to be studied   at LISA frequencies above their own confusion background.

Now consider the instrument-noise limited case, which will apply at high frequencies such as those observed with LIGO.  Again, since we require the power of the burst to exceed the noise power in the time interval $f^{-1}$, we can specify the instrument noise in equivalent power units, $\Omega_I$, the intensity of an isotropic stochastic background that would match the instrument noise during the integration interval $\approx f^{-1}$. 
Similarly to the previous case we write
\begin{equation}
\tau_{I>H>P}=[\Omega_{GH}/\Omega_I]^{-3/2}\tau_{P<H}
\end{equation}
which becomes 
\begin{equation}
\tau_{I>H>P}\approx  10^{12} \ {\rm sec} \ SNR_5^3 (\Omega_Ih^2/10^{-3})^{3/2}
(G\mu/10^{-12})^{-11/6}p^1(f/10^{2.5}{\rm Hz})^{-4/3}\gamma_{50}^{-11/4}\alpha_{0.1}^{-1/2},
\end{equation}
where we have normalized to typical $f$ and (very roughly) $\Omega_I$ for Advanced LIGO\cite{Cutler:2002me}. (We have scaled   from the multi-year sensitivity level of $\Omega_I\approx 0.5\times 10^{-8}$, since in $\Omega_I$ units the noise is proportional to the inverse square root of integration time\cite{Hogan:2001jn}).  This estimate suggests that even for large loops, bursts will be observable with Advanced LIGO only for $G\mu\approx 10^{-10}$ or larger, in rough agreement with the more detailed analysis of  Siemens et al. \cite{Siemens:2006vk}. As we have just seen, the prospects for detecting bursts are somewhat better with LISA. 
 
Within the generous errors we have allowed for burst rates in  this exploratory discussion, the   rate estimated here agrees approximately with\cite{Damour:2004kw} and  \cite{Siemens:2006vk}  if we evaluate at the value of $\alpha$ for which loops are forming at $t_{eq}$ and allow for different cosmological and filter-threshold assumptions.    The general conclusion is that the stochastic background, and not bursts, is the the best way to find and constrain light strings, but there is a still a window where bursts might be found and provide a valuable signature of string activity.

\section{detection sensitivity and noise sources}

The primary detection techniques under consideration for measuring the stochastic background are timing of millisecond pulsars,\cite{Kaspi:1994hp,Thorsett:1996dr,McHugh:1996hd,Lommen:2001ax,Lommen:2002je}, which is sensitive to gravitational waves with frequencies of order the inverse observation time (with the greatest sensitivity at longest periods, of order  tens of years), and the spaceborne Laser Interferometer Space Antenna (LISA), currently under development, which is sensitive to waves in a broad band  from $10^{-2}$ to $10^{-4}$ Hz.\cite{Hogan:2001jn} 

Apart from the burst detection, ground-based interferometers are not competitive; even advanced LIGO will at best reach a level\cite{Cutler:2002me} $\Omega_Gh^2\approx 0.5\times 10^{-8}$, and will not be the best tool to detect stochastic backgrounds from  very light strings. 

Figure 1 shows an estimate of sensitivity from a current pulsar study; the data have not yet been converted into a formal limit, but this gives an estimate of the currently attainable sensitivity level using the Pulsar Timing Array (PTA). 
A current benchmark of MSP sensitivity is the maximal monochromatic  fit by Lommen\cite{Lommen:2002je} to residuals in the cleanest pulsar data, PSR B1855+09, which yielded $\Omega_Gh^2<2\times10^{-9}$ at $f=1.9\times 10^{-9}$Hz. This corresponds to the maximum density of a  gravitational wave of that single frequency allowed by the data, assuming the wave is oriented optimally to produce the largest possible signal in the direction of PSR B1855+09.  Although this is not  yet presented as a rigorous upper limit it is shown in Fig. 1 as a guide to the sensitivity level possible with current timing accuracy and pulsar behavior.

The view adopted here is that the actual limit from the best current data will be within a factor of a few of this estimate. This view contrasts with that of Damour and Vilenkin\cite{Damour:2004kw}, who drew attention to conspicuous noise in some of the plotted data stream from this pulsar, and adopted an interpretation of the complete dataset that would allow significantly higher backgrounds. In this situation however, the noisy data came from one telescope system (the  Green Bank Telescope) and much more noise-free data from another (Arecibo);  the interpretation adopted here is the natural one based on only data from  the best understood  (and lowest noise) system.

The sensitivity curve is also shown for LISA in two ways: the upper curve is the sensitivity to a stochastic background at the maximum frequency resolution (of the order of the inverse observation interval); the lower curve is the sensitivity, after three years, including  information from a broad band,  assuming that stronger confusion backgrounds are resolved into sources,  and using the symmetric Sagnac combination\cite{armstrong,tintoarmstrong,tinto} as a calibration of non-gravitational wave backgrounds.  The Sagnac technique can use information from a broad band to measure a  broad band background, and thereby gains up to two   orders of magnitude in sensitivity for LISA to broadband stochastic backgrounds, depending on frequency\cite{Hogan:2001jn}.  Depending on details of the mission still under development, the actual sensitivity below $10^{-4}$ Hz may be significantly worse than shown.

In addition to the instrumental noise, the detection of backgrounds  is limited by astrophysical confusion noise. This ultimately limits the sensitivity of PTA, and of  LISA at low frequencies. 

Consider for illustration a binary system losing its binding energy predominantly to gravitational radiation.  It radiates,  over a decay time, the binding energy per mass $E  \approx 2GM/R$ in a broad band at the orbital frequency, $f \approx (M/R^3)^{1/2}$. Thus a homogeneous population of binaries  over time creates a background\cite{Phinney:2001di} with $\Omega_G\propto E\propto f^{2/3}$.  At low frequencies the background is confused  because of a very large number of sources; at higher frequencies, the sources are rare and more intense, so that an experiment can separate out the individual sources by resolving them in frequency. The curves in Fig. 1 terminate at high $f$ at around the frequency where the confusion from these backgrounds will be resolved out after a three year mission.

 Backgrounds from known populations of binaries  shown in Fig. 1 in the LISA band are taken from\cite{Hogan:2001jn}; for convenience, the Appendix recapitulates assumptions and caveats in those estimates. At lower frequencies, the confusion background from massive black hole binaries\cite{Rajagopal:1994zj,Jaffe:2002rt,Wyithe:2002ep,Enoki:2004ew} will ultimately limit the sensitivity of any pulsar experiment; indeed this  background may be detected 
in the near future using the improved Pulsar Timing Array. That population of massive binaries will be resolved at higher frequency by LISA, and  individual  binaries will be detected with very high signal-to-noise ratio.  In the lower part of the LISA band, confused signals from various populations of Galactic binary stars   dominate the noise. In the upper part of the band, the Galactic binaries can be resolved out, but   at these higher frequencies an irreducible background remains from white dwarf and other compact object binaries in other galaxies. At still higher frequencies of the order of 0.1 to 1 Hz, above the LISA band but accessible in principle to a similar  mission with shorter arms\cite{Seto:2005qy,Crowder:2005nr}, the white dwarf binaries disappear to mergers, and even the extragalactic  neutron star and black hole backgrounds might eventually be resolved out. It could be that this band would allow many orders of magnitude improvement in $\Omega_G$ sensitivity (although the possibility remains of other astrophysical noise from faint numerous sources that would limit the measurement.)

It is worth noting that,  if the string background is detected, it will only  be possible to separate the very numerous loop sources in the rare bursts, so the stochastic loop background will be the dominant source of noise for LISA at whatever frequencies it is detected.

\section{current and future detectability}

Unless otherwise noted, for simplicity the following comments are made for $\alpha=0.1, p=1$, as plotted in Fig. 1. At fixed $f$, the  predicted loop radiation signal scales in this model in the  three frequency regimes like:    $\mu^{1/2}\alpha^{1/2}p^{-1}$ (high $f$, H population dominated);
$\mu^{1/6}\alpha^{1/2}p^{-1}$ (intermediate $f$, P population dominated); $\mu^2\alpha^{1/2}p^{-1}$  (low $f$, P population dominated). These can be used to generalize to other combinations of $\mu,\alpha,p$. 

 The current  data on PSR B1855+09, with $\Omega_Gh^2<2\times10^{-9}$ at $f=1.9\times 10^{-9}$Hz, appear  to  set an upper limit  on  $G\mu$ of about $10^{-10}$: at this value, the predicted flux lies about an order of magnitude above the monochromatic flux limit\cite{Lommen:2002je}. At $G\mu=10^{-10}$, the pulsar bound comes from close to (or just above) $f_{peak}$, so the background is truly stochastic, with no detectable burst events expected.   
 Note  that at $G\mu=10^{-10}$ and above, the predicted flux at this frequency scales like $\Omega_G\propto (G\mu)^{1/6}\alpha^{1/2}p^{-1}$, so the limit on $G\mu$ is quite sensitive to factors of order unity in the precise value of the limit on $\Omega_G$. The limit also gets weaker for smaller $\alpha$.  If it is true, as assumed here, that $\alpha\approx 0.1$, this limit is already in the middle of the range expected in  brane inflation models, so a more careful evaluation of the bound is well motivated. That analysis should include a reasonable prior expectation of the   predicted spectrum, which in this regime is not flat (and includes,  in particular,   features  near $f_{peak}$ due to recent cosmic acceleration.)  
 
 At this level, bursts from the P loops may be detectable  with Advanced LIGO data\cite{Siemens:2006vk}. Such detections would serve both to confirm the stringy nature of the background source, and to break degeneracies in the parameters by separating H and P populations.

Although  it appears   that a value of $G\mu$ significantly above $10^{-10}$ is already ruled out,  a different behavior sets in just below $G\mu=10^{-10}$: since $f_{peak}$ for $G\mu< 10^{-10.3}$ lies above the frequency where the pulsars are most sensitive, the turndown in the loop spectrum at $f<f_{peak}$ makes it significantly  more difficult to set limits below $G\mu\approx 10^{-10.3}$. Thus, the current limit probably is within a factor of a few of $G\mu=10^{-10}$.
 Future projects will achieve better sensitivity by improving timing residuals and increasing the number of pulsars. They will eventually be limited however by the MBHB background,  which will limit sensitivity for significant detection or upper limits  to not better than about $G\mu\approx 10^{-11}$.   
 
The smallest $G\mu$ reachable by LISA appears to be in the Sagnac mode in a window around $10^{-4}$ Hz. At $G\mu\approx 10^{-15}$ the background lies about a factor of three above the binary confusion flux, which in turn is about twice the broadband (Sagnac) instrument limit.  Again, this is close to $f_{peak}$ so it is stochastic and dominated by the P population, with little opportunity to see bursts.

For higher $G\mu$ than this, the loop background is a significant additional noise source for LISA.  Although the  P radiation dominates the total power in part  of the LISA band for $10^{-15}<G\mu<10^{-13}$, as we have seen  the individual burst events are seldom detectable as they are hidden in the confusion noise. For larger $G\mu$ than about $ 10^{-12}$, the confusion noise across the entire main LISA band--- about a factor of 100 in frequency--- is dominated by stochastic H  population radiation that is the first and most conspicuously detectable effect of the strings. 
For  $G\mu$ greater  than about $ 10^{-11}$,  LISA-detectable bursts above this noise  can occur often enough that their  statistics can be used as a diagnostic of string properties. 

These results confirm the generally robust prediction of burst events from strings
\cite{Damour:2000wa,Damour:2001bk,Damour:2004kw,Siemens:2006vk} but also show that for small $G\mu$, the mean stochastic background is a much more sensitive probe than bursts.

It is possible that confusion backgrounds are small enough to  allow detection  for $G\mu<10^{-20}$ if a suitable detector could be built at 0.1 to 1 Hz , with sensitivity enough to resolve all the extragalactic binary sources and reach an instrument noise limit on the order of $\Omega_Gh^2\approx 10^{-15}$, the level required  to search for inflationary graviton perturbations
at zero spectral tilt (see e.g., \cite{Seto:2005qy,Crowder:2005nr}).  
(It is of course also possible that unresolvable backgrounds not yet modeled could make this level unreachable.) By the same token, strings would have to be at least as light as this to avoid introducing a contaminating foreground that would make the inflationary waves undetectable.

\begin{acknowledgements}
This work was supported by NSF grant AST-0098557 at the University of
Washington. 
\end{acknowledgements}

\break
\section{APPENDIX: SENSITIVITY LIMITS AND BINARY BACKGROUNDS } 
This appendix  summarizes for convenience the backgrounds  shown in Fig. 1.
More detail on the  arguments for these numbers, and on the maximally sensitive Sagnac techniques for measuring the stochastic background with LISA-type interferometers,  can be found  in \cite{Hogan:2001jn}.

The approximate threshold sensitivity of the planned LISA antenna with $5\times 10^6$ km arm
lengths and for a signal-to-noise ratio S/N = 1 is shown in Figure 1.  The
sensitivity using 
the standard Michelson observable    can
be
approximated by a set of power law segments:
\begin{eqnarray}
h_{rms } & = & 1.0\times 10^{-20} [f/10 mHz]/\sqrt{\rm Hz},\qquad  10 mHz < f \\ \nonumber
& & 1.0\times 10^{-20}/\sqrt{\rm Hz},\qquad 2.8 mHz < f < 10 mHz\\ \nonumber
& & 7.8\times 10^{-18} [(0.1 mHz/f)^2]/\sqrt{\rm Hz}, \qquad 0.1 mHz < f < 2.8mHz\\ \nonumber
& & 7.8\times 10^{-18} [(0.1 mHz/f)^{2.5}]/\sqrt{\rm Hz}, \qquad 0.01 mHz < f<0.1mHz   
\end{eqnarray}
where   the sensitivity has been averaged over the
source
directions.  Below 100 $\mu$Hz there is no adopted mission sensitivity
requirement, but the listed sensitivity might be achieved for frequencies down to (say) 10 $\mu$Hz. More accurate modeling can be found in\cite{Prince:2002hp,Krolak:2004xp}.

Experiments characterize backgrounds by  $h_{rms}^2$, the spectral density of the
gravitational wave strain (also sometimes denoted $S_h$).  For cosmology, we are interested 
in sensitivity in terms of the broadband energy density of an isotropic, unpolarized,
stationary background, whose cosmological importance is characterized by
\begin{equation}
\Omega_{GW}(f)\equiv  
\rho_c^{-1} {d\rho_{GW}\over d \ln f}
={4\pi^2\over 3 H_0^2}f^3 h_{rms}^2(f)
\end{equation}
where we   adopt  units of the critical density $\rho_c$.
The  broadband energy density per $e$-folding of frequency, $\Omega_{GW}(f)$, is
thus related to the   rms strain spectral density 
 by\cite{maggiore}
\begin{equation}
{h_0^2\Omega_{GW}\over 10^{-8}}
\approx
\left({h_{rms}(f )\over 2.82\times  10^{-18}{\rm \ Hz^{-1/2}}}\right)^2
\left({f\over 1{\rm mHz}}\right)^3,
\end{equation}
where $h_0$ conventionally denotes Hubble's constant in units of 
$100{\rm km\ s^{-1}\ Mpc^{-1}}$.

Figure  1 shows the
  effective  confusion noise from both galactic and extragalactic
binaries remaining
after the resolved binaries have been fitted out of the data record 
(see e.g. \cite{benderhils,Farmer:2003pa}).  Essentially
none of the extragalactic stellar-mass
binaries can be resolved with LISA's sensitivity\cite{Farmer:2003pa} (in contrast to intense signals from 
an expected small number involving massive black hole binaries).
There is about a factor of three uncertainty in the estimated total galactic signal
level  (which translates to about an order of magnitude in $\Omega$ units), and more in 
  the estimated extragalactic signal level.

The
normalization for various populations of galactic binaries (UB+WUMa+GCWDB+CV) is taken from 
levels estimated in \cite{hils90} for the total binary
backgrounds,  with estimates from \cite{benderhils} for
  the reduction of confusion noise at higher frequencies
by fitting out Galactic binaries.  
For close white dwarf
binaries (CWDBs)\cite{webbink,hils90}, this value  lies within a factor 2 of \cite{webbink2}.  The
GCWDBs\cite{hils90,benderhils,hilsbender,webbink,webbink2,kosenko,nelemans,schneider,nelemans2}
include He-He, He-CO and CO-CO white dwarf binaries, as well as a few binaries containing the
rarer O/Ne/Mg white dwarfs.

The extragalactic XGCWDB background 
goes away above about 0.1 Hz, provided that merger-phase and ringdown radiation
from coalescences are not significant.  At higher frequencies, the binary
background is expected to be almost entirely due to extragalactic neutron
star
binaries and 5 to 10 solar mass black hole
binaries. 
The neutron star binary coalescence rate in our galaxy is taken as $1\times 10^{-5} {\rm
yr^{-1}}$.  This
estimate  has a high uncertainty \cite{kalogera}, but gives a plausible estimate of the total gravitational wave
background level, allowing for some additional contribution from black hole
binaries. A small correction is included  for eccentricity of the NS-NS binaries\cite{hils91}, and we adopt the ratio of
0.3 between the extragalactic and galactic amplitudes\cite{kosenko}.
  The   BH-BH binaries may be the dominant source\cite{schutz,kalogera}.  

\begin{thebibliography}{}
\bibitem{kibble}
T. W. B. Kibble, J.  Phys. A 9, 1387 (1976)
\bibitem{Zeldovich:1980gh}
Y.~B. Zeldovich, Mon. Not. Roy. Astron. Soc. {\bf 192},  663  (1980).

\bibitem{Vilenkin:1981iu}
A. Vilenkin, Phys. Rev. Lett. {\bf 46},  1169  (1981).

\bibitem{Vilenkin1981}
A. Vilenkin, Phys. Lett. {\bf 107B}, 47 (1981)

 
  \bibitem{Hindmarsh:1994re}
M.~B. Hindmarsh and T.~W.~B. Kibble, Rept. Prog. Phys. {\bf 58},  477  (1995),
  hep-ph/9411342.

\bibitem{vilenkinshellard}
A. Vilenkin, E. P. S. Shellard, {\it Cosmic Strings and Other Topological Defects}, Cambridge University Press (2000)

\bibitem{Jones:2002cv}
N. Jones, H. Stoica, and S.~H.~H. Tye, JHEP {\bf 07},  051  (2002),
  hep-th/0203163.

\bibitem{Sarangi:2002yt}
S. Sarangi and S.~H.~H. Tye, Phys. Lett. {\bf B536},  185  (2002),
  hep-th/0204074.
  \bibitem{Jones:2003da}
N.~T. Jones, H. Stoica, and S.~H.~H. Tye, Phys. Lett. {\bf B563},  6  (2003),
  hep-th/0303269.
\bibitem{Copeland:2003bj}
  E.~J.~Copeland, R.~C.~Myers and J.~Polchinski,
  JHEP {\bf 0406}, 013 (2004)
  [arXiv:hep-th/0312067].
  
   \bibitem{Jackson:2004zg}
M.~G. Jackson, N.~T. Jones, and J. Polchinski,   (2004), hep-th/0405229.


\bibitem{polchinski}
  J.~Polchinski, 2004 Cargese Lectures,
  arXiv:hep-th/0412244 (2004)

\bibitem{Polchinski:2004hb}
J.~Polchinski,
arXiv:hep-th/0410082.

\bibitem{davis}
A. C. Davis, T. W. B. Kibble, Contemp. Phys. 46, 313 (2005)
\bibitem{kibble2}
T.~W.~B.~Kibble,
  arXiv:astro-ph/0410073.
\bibitem{vilenkin2}
 A.~Vilenkin,
  arXiv:hep-th/0508135.
  
 
\bibitem{Kachru:2003sx}
  S.~Kachru, R.~Kallosh, A.~Linde, J.~Maldacena, L.~McAllister and S.~P.~Trivedi,
  JCAP {\bf 0310}, 013 (2003)
  [arXiv:hep-th/0308055].
  
   

\bibitem{Hogan:1984is}
  C.~J.~Hogan and M.~J.~Rees,
  Nature {\bf 311}, 109 (1984).

\bibitem{Vachaspati:1986cc}
T. Vachaspati and A. Vilenkin, Phys. Rev. {\bf D35},  1131  (1987).


  
 \bibitem{Bennett:1989yp}
D.~P.~Bennett and F.~R.~Bouchet,
Phys.\ Rev.\ D {\bf 41}, 2408 (1990);


\bibitem{Caldwell:1991jj}
  R.~R.~Caldwell and B.~Allen,
  Phys.\ Rev.\ D {\bf 45}, 3447 (1992).

\bibitem{Caldwell:1996en}
  R.~R.~Caldwell, R.~A.~Battye and E.~P.~S.~Shellard,
  Phys.\ Rev.\ D {\bf 54}, 7146 (1996)
  [arXiv:astro-ph/9607130].
  
\bibitem{Allen:1990tv}
  B.~Allen and E.~P.~S.~Shellard,
  Phys.\ Rev.\ Lett.\  {\bf 64}, 119 (1990).

  \bibitem{Austin:1993rg}
D. Austin, E.~J. Copeland, and T.~W.~B. Kibble, Phys. Rev. {\bf D48},  5594
  (1993), hep-ph/9307325.


\bibitem{Siemens:2002dj}
X.~Siemens, K.~D.~Olum and A.~Vilenkin,
Phys.\ Rev.\ D {\bf 66}, 043501 (2002)
[arXiv:gr-qc/0203006].

\bibitem{Siemens:2003ra}
X. Siemens and K.~D. Olum, Phys. Rev. {\bf D68},  085017  (2003),
  gr-qc/0307113.


\bibitem{Vanchurin:2005pa}
  V.~Vanchurin, K.~D.~Olum and A.~Vilenkin,
  arXiv:gr-qc/0511159.
  
\bibitem{Ringeval:2005kr}
  C.~Ringeval, M.~Sakellariadou and F.~Bouchet,
  arXiv:astro-ph/0511646.
  
\bibitem{Martins:2005es}
  C.~J.~A.~Martins and E.~P.~S.~Shellard,
  Phys.\ Rev.\ D {\bf 73}, 043515 (2006)
  [arXiv:astro-ph/0511792].

   
\bibitem{Sakellariadou:2004wq}
  M.~Sakellariadou,
  JCAP {\bf 0504}, 003 (2005)
  [arXiv:hep-th/0410234].

\bibitem{Avgoustidis:2005nv}
  A.~Avgoustidis and E.~P.~S.~Shellard,
  Phys.\ Rev.\ D {\bf 73}, 041301 (2006)
  [arXiv:astro-ph/0512582].


\bibitem{Damour:2000wa}
T. Damour and A. Vilenkin, Phys. Rev. Lett. {\bf 85},  3761  (2000),
  gr-qc/0004075.

\bibitem{Damour:2001bk}
T. Damour and A. Vilenkin, Phys. Rev. {\bf D64},  064008  (2001),
  gr-qc/0104026.

\bibitem{Damour:2004kw}
  T.~Damour and A.~Vilenkin,
   ``Gravitational radiation from cosmic (super)strings: Bursts, stochastic
  Phys.\ Rev.\ D {\bf 71}, 063510 (2005)
  [arXiv:hep-th/0410222].
  
\bibitem{Siemens:2006vk}
  X.~Siemens, J.~Creighton, I.~Maor, S.~R.~Majumder, K.~Cannon and J.~Read,
  arXiv:gr-qc/0603115.

\bibitem{Hogan:2001jn}
  C.~J.~Hogan and P.~L.~Bender,
  Phys.\ Rev.\ D {\bf 64}, 062002 (2001)
  [arXiv:astro-ph/
  ].

  
\bibitem{Turok:1984cn}
N. Turok, Nucl. Phys. {\bf B242},  520  (1984).


\bibitem{Cutler:2002me}
  C.~Cutler and K.~S.~Thorne,
  arXiv:gr-qc/0204090.


\bibitem{Kaspi:1994hp}
V.~M.~Kaspi, J.~H.~Taylor and M.~F.~Ryba,
Astrophys.\ J.\  {\bf 428}, 713 (1994).

\bibitem{Thorsett:1996dr}
S.~E.~Thorsett and R.~J.~Dewey,
Phys.\ Rev.\ D {\bf 53}, 3468 (1996).

\bibitem{McHugh:1996hd}
  M.~P.~McHugh, G.~Zalamansky, F.~Vernotte and E.~Lantz,
  Phys.\ Rev.\ D {\bf 54}, 5993 (1996).



\bibitem{Lommen:2001ax}
  A.~N.~Lommen and D.~C.~Backer,
  Astrophys.\ J.\  {\bf 562}, 297 (2001)
  [arXiv:astro-ph/0107470].

\bibitem{Lommen:2002je}
A.~N. Lommen, astro-ph/0208572.

   \bibitem{armstrong}
J. W. Armstrong,  F. B. Estabrook, 
\& M. Tinto,   ApJ 527, 814  (1999) 
\bibitem{tintoarmstrong}
M. Tinto  \& J. Armstrong, Phys. Rev. D 59, 102003 (1999)
\bibitem{estabrook}
 F. B. Estabrook,  M. Tinto 
\&  J. W. Armstrong, Phys. Rev. D 62, 042002  (2000)
\bibitem{tinto}
 M. Tinto, J. W. Armstrong  \& F. B. Estabrook, Phys. Rev. D, 63, 021101 (R)  (2000)

  

\bibitem{Phinney:2001di}
  E.~S.~Phinney,
  arXiv:astro-ph/0108028.

\bibitem{Rajagopal:1994zj}
  M.~Rajagopal and R.~W.~Romani,
  Astrophys.\ J.\  {\bf 446}, 543 (1995)
  [arXiv:astro-ph/9412038].
\bibitem{Jaffe:2002rt}
  A.~H.~Jaffe and D.~C.~Backer,
  Astrophys.\ J.\  {\bf 583}, 616 (2003)
  [arXiv:astro-ph/0210148].

\bibitem{Wyithe:2002ep}
  J.~S.~B.~Wyithe and A.~Loeb,
  Astrophys.\ J.\  {\bf 590}, 691 (2003)
  [arXiv:astro-ph/0211556].

 \bibitem{Enoki:2004ew}
  M.~Enoki, K.~T.~Inoue, M.~Nagashima and N.~Sugiyama,
  Astrophys.\ J.\  {\bf 615}, 19 (2004)
  [arXiv:astro-ph/0404389].
  
\bibitem{Seto:2005qy}
  N.~Seto,
  Phys.\ Rev.\ D {\bf 73}, 063001 (2006)
  [arXiv:gr-qc/0510067].
  
\bibitem{Crowder:2005nr}
  J.~Crowder and N.~J.~Cornish,
  Phys.\ Rev.\ D {\bf 72}, 083005 (2005)
  [arXiv:gr-qc/0506015].

\bibitem{Prince:2002hp}
  T.~A.~Prince, M.~Tinto, .~L.~Larson and .~W.~Armstrong,
  Phys.\ Rev.\ D {\bf 66}, 122002 (2002)
  [arXiv:gr-qc/0209039].
  
\bibitem{Krolak:2004xp}
  A.~Krolak, M.~Tinto and M.~Vallisneri,
  Phys.\ Rev.\ D {\bf 70}, 022003 (2004)
  [arXiv:gr-qc/0401108].

  
  \bibitem{maggiore}
 M. Maggiore,  Phys. Rep. 331, 283 (2000)
\bibitem{benderhils}
P. L.  Bender \&  D. Hils, Class. Quant. Grav. 14, 1439  (1997)

\bibitem{Farmer:2003pa}
  A.~J.~Farmer and E.~S.~Phinney,
  Mon.\ Not.\ Roy.\ Astron.\ Soc.\  {\bf 346}, 1197 (2003)
  [arXiv:astro-ph/0304393].
 
  \bibitem{hils90}
D. Hils, P. L. Bender,  \&  R. F. Webbink, ApJ 360, 75 (1990)
\bibitem{webbink}
R. F. Webbink, Astrophys. J. 277, 355  (1984)
\bibitem{webbink2}
R. F. Webbink  \& Z. Han , in ``Laser Interferometer Space
Antenna:
  Proc. 2nd Int. LISA Symp.",  p. 61  (1998)
\bibitem{kosenko}
 D. I. Kosenko \& K. A. Postnov, Astron. Astrophys. 336, 786 (1998)
 \bibitem{hilsbender}
D. Hils  \&  P. L. Bender, ApJ 53, 334  (2000)

\bibitem{nelemans}
G. Nelemans, L. R. Yungelson, S. F. Portegies Zwart,  \&  F. Verbunt,
Astron. Astrophys. 365, 491 (2001)
\bibitem{schneider}
 R. Schneider,  V. Ferrari, S.  Matarrese \&  S.
F. Portegies Zwart,
  Mon. Not. R. Astron. Soc., submitted, astro-ph/0002055  (2000)
\bibitem{nelemans2}
 G. Nelemans,  F. Verbunt, L. R. Yungelson  \& S.
F. Portegies Zwart, Astron. Astrophys., 360, 1011  (2000)


\bibitem{kalogera}
V. Kalogera, in {\it Gravitational Waves: Third
Edoardo Amaldi Conference}, ed. S. Meshkov (AIP Conf. Proc. Vol. CP523), 41 (2000)
\bibitem{hils91}
D.  Hils, Astrophys. J. 381, 484  (1991)

\bibitem{schutz}
 B. F. Schutz, Class. Quant. Grav. 16, A131 (1999)


\end{thebibliography}
\end{document}